\documentclass[prd,eqsecnum,twocolumn,showpacs,preprintnumbers,amsmath,amssymb]{revtex4}

\usepackage{graphicx}

\usepackage{bm}

\newcommand{\beq}{\begin{equation}}
\newcommand{\eeq}{\end{equation}}
\newcommand{\beqs}{\begin{eqnarray}}
\newcommand{\eeqs}{\end{eqnarray}}

\begin{document}

\title{Lower Bounds on the Ground State Entropy of the Potts Antiferromagnet
 on Slabs of the Simple Cubic Lattice} 

\author{Robert Shrock and Yan Xu}

\affiliation{
C. N. Yang Institute for Theoretical Physics \\
State University of New York \\
Stony Brook, NY 11794}

\begin{abstract}

  We calculate rigorous lower bounds for the ground state degeneracy per site,
$W$, of the $q$-state Potts antiferromagnet on slabs of the simple cubic
lattice that are infinite in two directions and finite in the third and that
thus interpolate between the square (sq) and simple cubic (sc) lattices. We
give a comparison with large-$q$ series expansions for the sq and sc lattices
and also present numerical comparisons.

\end{abstract}


\maketitle

\section{Introduction} 

   Nonzero ground state entropy (per lattice site), $S_0 \ne 0$, is an
important subject in statistical mechanics, as an exception to the third law of
thermodynamics and a phenomenon involving large disorder even at zero
temperature. Since $S_0 = k_B \ln W$, where $W = \lim_{n \to \infty}
W_{tot.}^{1/n}$ and $n$ denotes the number of lattice sites, $S_0 \ne 0$ is
equivalent to $W > 1$, i.e., a total ground state degeneracy $W_{tot.}$ that
grows exponentially rapidly as a function of $n$. One physical example is
provided by H$_2$O ice, for which the residual entropy per site (at 1
atm. pressure) is measured to be $S_0 = (0.41 \pm 0.03)k_B$, or equivalently,
$W=1.51 \pm 0.05$ \cite{pauling35}-\cite{berg07}.  A salient property of ice is
that the ground state entropy occurs without frustration; i.e., each of the
ground state configurations of the hydrogen atoms on the bonds between oxygen
atoms minimizes the internal energy of the crystal \cite{frus}.  

  A model that also exhibits ground state entropy without frustration and hence
provides a useful framework in which to study the properties of this phenomenon
is the $q$-state Potts antiferromagnet \cite{wurev}-\cite{chowwu} on a given
lattice $\Lambda$ or, more generally, a graph $G$, for sufficiently large $q$.
Consider a graph $G=(V,E)$, defined by its vertex (site) and edge (bond) sets
$V$ and $E$.  Denote the cardinalities of these sets as $n(G) = |V| \equiv n$
and $e(G)=|E|$, and let $\{G\} \equiv \lim_{n(G) \to \infty} G$. An important
connection with graph theory is the fact that the zero-temperature partition
function of the $q$-state Potts antiferromagnet on the graph $G$ satisfies
$Z(G,q,T=0)=P(G,q)$, where $P(G,q)$ is the chromatic polynomial expressing the
number of ways of coloring the vertices of $G$ with $q$ colors such that no two
adjacent vertices have the same color (called a proper $q$-coloring of $G$)
\cite{newton,jemrev}.  Thus,
\beq 
W(\{G\},q) = \lim_{n \to \infty}P(G,q)^{1/n} \ . 
\label{w}
\eeq
In general, for certain special values of $q$, denoted $q_s$, one has the
following noncommutativity of limits \cite{w} 
\beq
\lim_{n \to \infty} \lim_{q \to q_s} P(G,q)^{1/n} \ne 
\lim_{q \to q_s} \lim_{n \to \infty} P(G,q)^{1/n} \ , 
\label{wnoncom}
\eeq
and hence it is necessary to specify which order of limits that one takes in
defining $W(\{G\},q)$.  Here by $W(\{G\},q)$ we mean the function obtained by
setting $q$ to the given value first and then taking $n \to \infty$.  For the
$n \to \infty$ limit of a bipartite graph $G_{bip.}$, an elementary lower bound
is $W(\{G_{bip.} \},q) \ge \sqrt{q-1}$, so that for $q > 2$, the Potts
antiferromagnet has a nonzero ground state entropy on such a lattice.  A better
lower bound for the square lattice is $W(sq,q) \ge (q^2-3q+3)/(q-1)$
\cite{biggs77}.  In previous work \cite{ww}-\cite{312} one of us and Tsai
derived lower and upper bounds on $W$ for a variety of different
two-dimensional lattices.  It was found that these lower bounds are quite close
to the actual values as determined with reasonably good accuracy from large-$q$
series expansions and/or Monte Carlo measurements. 

In the present paper we generalize these lower bounds on two-dimensional
lattice graphs by deriving lower bounds on $W(\{G\},q)$ for sections of a
three-dimensional lattice, namely the simple cubic lattice, which are of
infinite extent in two directions (taken to lie along the $x$ and $y$ axes) and
finite in the third direction, $z$.  By comparison with large-$q$ expansions
and numerical evaluations, we show how the lower bounds for the $W$ functions
for these slabs interpolate between the values for the (respective
thermodynamic limits of the) square and simple cubic lattices.  These bounds
are of interest partly because one does not know the exact functions $W(sq,q)$
or $W(sc,q)$ for general $q$.

\section{Calculational Method}

    Let us consider a section (slab) of the simple cubic lattice of dimensions
$L_x \times L_y \times L_z$ vertices, which we denote $sc[(L_x)_{BCx} \times
(L_y)_{BCy} \times (L_z)_{BCz}]$, where the boundary conditions (BC) in each
direction are indicated by the subscripts.  The chromatic polynomial of this
lattice will be denoted $P(sc[(L_x)_{BCx} \times (L_y)_{BCy} \times
(L_z)_{BCy}],q)$.  We will calculate lower bounds for $W(sc[(L_x)_{BCx} \times
(L_y)_{BCy} \times (L_z)_{BCz}],q)$ in the limit $L_x \to \infty$ and $L_y \to
\infty$ with $L_z$ fixed.  These are independent of the boundary conditions
imposed in the directions in which the slab is of infinite extent, and hence,
for brevity of notation, we will denote the limit $\lim_{L_x, \ L_y \to
\infty} sc[(L_x)_{BCx} \times (L_y)_{BCy} \times (L_z)_{BCz}]$ simply as
$S_{(L_z)_{BC_z}}$, where $S$ stands for ``slab''. We will consider both free
(F) and periodic (P) boundary conditions in the $z$ direction, and thus slabs
such as $S_{3_F}$, $S_{3_P}$, etc.  For technical reasons (to get an expression
involving a trace of a coloring matrix, as explained below) we will use
periodic boundary conditions in the $x$ direction. Note that the proper
$q$-coloring constraint implies that FBC$_z$ and PBC$_z$ are equivalent if
$L_z=2$. The number of vertices for $G=sc[(L_x)_{BCx} \times (L_y)_{BCy} \times
(L_z)_{BCz}]$ is $n=L_xL_yL_z$.  The specific form of Eq. (\ref{w}) for our
calculation is
\begin{widetext}
\beq
W(S_{(L_z)_{BCz}},q) = \lim_{L_y \to \infty} \lim_{L_x \to \infty}
[P(sc[(L_x)_P \times (L_y)_{BCy} \times (L_z)_{BCz}],q)]^{1/n} \ . 
\label{wform}
\eeq
\end{widetext}

To derive a lower bound on $W(S_{(L_z)_{BC_z}},q)$, we generalize the method of
Refs. \cite{biggs77}-\cite{wn} from two to three dimensions.  We consider two
adjacent transverse slices of the slab orthogonal to the $x$ direction, with
$x$ values $x_0$ and $x_0+1$.  These are thus sections of the square lattice of
dimension $L_y \times L_z$, which we denote $G_{x_0}=sq[(L_y)_{BC_y} \times
(L_z)_{BC_z}]_{x_0}$ and $G_{x_0+1} = sq[(L_y)_{BC_y} \times
(L_z)_{BC_z}]_{x_0+1}$.  We label a particular color assignment to the vertices
of $G_{x_0}$ that is a proper $q$-coloring of these vertices as $C(G_{x_0})$
and similarly for $G_{x_0+1}$. The total number of proper $q$-colorings of
$G_{x_0}$ is
\beq
{\cal N} = P(G_{x_0},q) = P(G_{x_0+1},q) \ .
\label{caln}
\eeq
Now let us add the edges in the $x$ direction that join these two adjacent
transverse slices of the slab together.  Among the ${\cal N}^2$ color
configurations that yield proper $q$-colorings of these two separate $yz$
transverse slices, some will continue to be proper $q$-colorings after we add
these edges that join them in the $x$ direction, while others will not.  We
define an ${\cal N} \times {\cal N}$-dimensional coloring compatibility matrix
$T$ with entries $T_{C(G_{x_0}),C(G_{x_0+1})}$ equal to (i) 1 if the color
assignments $C(G_{x_0})$ and $C(G_{x_0+1})$ are proper $q$-colorings after the
edges in the $x$ direction have been added joining $G_{x_0}$ and $G_{x_0+1}$,
i.e., if the color assigned to each vertex $v(x_0,y,z)$ in $G_{x_0}$ is
different from the color assigned to the vertex $v(x_0+1,y,z)$ in $G_{x_0+1}$;
and (ii) 0 if the color assignments $C(G_{x_0})$ and $C(G_{x_0+1})$ are not
proper $q$-colorings after the edges in the $x$ direction have been added,
i.e., there exists some color assigned to a vertex $v(x_0,y,z)$ in $G_{x_0}$
that is equal to a color assigned to the vertex $v(x_0+1,y,z)$ in $G_{x_0+1}$.
Clearly, $T_{ij} = T_{ji}$.  The chromatic polynomial for the slab is then
given by the trace
\beq
P(sc[(L_x)_P \times (L_y)_{BCy} \times (L_z)_{BC_z}],q) = {\rm Tr}(T^{L_x})
\ . 
\label{ptracex}
\eeq

Since $T$ is a real symmetric matrix, there exists an orthogonal matrix $A$
that diagonalizes $T$: $ATA^{-1} = T_{diag.}$.  Let us denote the ${\cal N}$
eigenvalues of $T$ as $\lambda_{T,j}$, $1 \le j \le {\cal N}$.  Since $T$ is a
real non-negative matrix, we can apply the generalized Perron-Frobenius theorem
\cite{lancaster85,minc88} to infer that $T$ has a real maximal
eigenvalue,
which we denote $\lambda_{T,max}$.  It follows that
\beq
\lim_{L_x \to \infty} [P(sc[(L_x)_P \times (L_y)_{BCy} \times (L_z)_{BC_z}],
q)]^{1/L_x} = \lambda_{T,max} \ . 
\label{tlambdamax}
\eeq
Now for the transverse slices $G_{x_0}$ and $G_{x_0+1}$, denoted generically
as $ts((L_z)_{BC_z})$, the chromatic polynomial has the form 
\beq
P(G_{x_0},q) = P(G_{x_0+1},q) = \sum_j c_j \, 
(\lambda_{ts((L_z)_{BC_z}),j})^{L_y}
\label{pgyz}
\eeq
where the $c_j$ are coefficients whose precise form is not needed here. 
The set of $\lambda_{ts((L_z)_{BC_z}),j}$'s is independent of the length $L_y$
and although this set depends on $BC_y$, the maximal one (having the largest
magnitude), $\lambda_{ts((L_z)_{BC_z}),max}$, is independent of $BC_y$ 
(e.g., \cite{s5} and references therein).  Hence, 
\beqs
\lim_{L_y \to \infty} [P(G_{x_0},q)]^{1/L_y} & \equiv & 
\lim_{L_y \to \infty} ({\cal N})^{1/L_y} \cr\cr
                 & = &  \lambda_{ts((L_z)_{BC_z}),max} \ . 
\label{tsdom}
\eeqs
The two adjacent slices together with the edges in the $x$ direction that join
them constitute the graph $sc[2_F \times (L_y)_{BCy} \times
(L_z)_{BC_z}]$. We denote the chromatic polynomial for this section (tube) of
the $sc$ lattice as $P(sc[2_F \times (L_y)_{BCy} \times (L_z)_{BC_z}],q)$
(which is equal to $P(sc[2_P \times (L_y)_{BCy} \times (L_z)_{BC_z}],q)$
because of the proper $q$-coloring condition). This has the form
\beqs
& & P(sc[2_F \times (L_y)_{BCy} \times (L_z)_{BC_z}],q) \cr\cr
& = & \sum_j c'_j \, (\lambda_{tube((L_z)_{BC_z}),j})^{L_y}
\label{ptube}
\eeqs
where $c'_j$ are coefficients analogous to those in (\ref{pgyz}). Therefore,
\beqs
\lim_{L_y \to \infty} 
& & [P(sc[2_F \times (L_y)_{BCy} \times (L_z)_{BC_z}],q)]^{1/L_y} =  \cr\cr
& = & \lambda_{tube((L_z)_{BC_z}),max} \ . 
\label{tubedom}
\eeqs

Now let us denote the column sum (CS) 
\beq
CS_j(T) = \sum_{i=1}^{\cal N} T_{ij} \ , 
\label{cs}
\eeq
which is equal to the row sum $\sum_{j=1}^{\cal N}T_{ij}$, since $T^T=T$. We
also define the sum of all entries (SE) of $T$ as
\beq
SE(T) = \sum_{i,j=1}^{\cal N} T_{ij} \ . 
\label{se}
\eeq
Note that $SE(T)/{\cal N}$ is the average row (= column) sum.  Next, we observe
that 
\beq
SE(T) = P(sc[2_F \times (L_y)_{BCy} \times (L_z)_{BC_z}],q) \ . 
\label{se_tube}
\eeq

 To obtain our lower bound, we then use the $r=1$ special case of the theorem
that for a non-negative symmetric matrix $T$ and $r \in {\mathbb N}_+$
\cite{london66}
\beq
\lambda_{max}(T) \ge \left [ \frac{SE(T^r)}{{\cal N}} \right ]^{1/r} \ . 
\label{london}
\eeq
The lower bound is then 
\beq
W(S_{(L_z)_{BC_z}},q) \ge W(S_{(L_z)_{BC_z}},q)_\ell 
\label{wlow}
\eeq
where
\begin{widetext}
\beqs
W(S_{(L_z)_{BC_z}},q)_\ell & = & \lim_{L_y \to \infty} 
\bigg ( \frac{SE(T)}{\cal N} \bigg )^{1/(L_yL_z)} \cr\cr
& & \cr\cr
& = & \lim_{L_y \to \infty} 
\bigg [ \frac{P(sc[2_F \times (L_y)_{BCy} \times (L_z)_{BC_z}],q)}
{P(sq[(L_y)_{BCy} \times (L_z)_{BC_z}],q)} \bigg ]^{1/(L_yL_z)} \cr\cr
& & \cr\cr
& = & \bigg [ \frac{\lambda_{tube((L_z)_{BC_z}),max}}
{\lambda_{ts((L_z)_{BC_z}),max}} \bigg  ]^{1/L_z}  \ . 
\label{wlowbound}
\eeqs
\end{widetext}

\section{Results for Slab of Thickness $L_z=2$ with FBC$_z$}

We now evaluate our general lower bound in Eqs. (\ref{wlow}) and
(\ref{wlowbound}) for a slab of the simple cubic lattice with 
thickness $L_z=2$ and FBC$_z$, denoted $S_{2_F}$.  In this case the
transverse slice is the graph $sq[2_F \times (L_y)_{BCy}]$. For FBC$_y$, 
an elementary calculation yields
\beq
P(sq[2_F \times (L_y)_F],q)=q(q-1)(q^2-3q+3)^{L_y-1}
\label{psq2f}
\eeq
with a single $\lambda_{ts(2_F)}=\lambda_{ts(2_P)} \equiv \lambda_{ts(2)}$, and
this is also the maximal $\lambda$ for PBC$_y$ \cite{bds,w}, so that
\beq
\lambda_{ts(2),max} = q^2-3q+3 \ . 
\label{lamts2}
\eeq
We next use the calculation of 
\beqs
P(sc[2_F \times (L_y)_F \times 2_F],q) & = & 
P(sc[2_F \times 2_F \times (L_y)_F],q) \cr\cr
& = & P(sq[4_P \times (L_y)_F],q) \cr\cr
& & 
\label{p22tube}
\eeqs
in Ref. \cite{hd} (where each of the $2_F$ BC's is equivalent to $2_P$), from
which we calculate the maximal $\lambda_{tube(2),max}$ to be 
\beq
\lambda_{tube(2),max}=\frac{1}{2}\bigg [ q^4-8q^3+29q^2-55q+46 + \sqrt{R_{22}}
\ \bigg ]
\label{lamtubemax}
\eeq
where
\beqs
R_{22} & = & q^8-16q^7+118q^6-526q^5+1569q^4 \cr\cr
  & - & 3250q^3+4617q^2-4136q+1776 \ . 
\label{r}
\eeqs
We then substitute these results for $\lambda_{ts(2),max}$ and
$\lambda_{tube(2),max}$ into the $L_z=2$ special case of (\ref{wlowbound}) to
obtain $W(S_2,q)_{\ell}$, and thus the resultant lower bound on
$W(S_{2_F},q)=W(S_{2_P},q) \equiv W(S_2,q)$: 
$W(S_2,q) \ge W(S_2,q)_\ell$.

\section{Comparison with Large-$q$ Series Expansions}

One way to elucidate how this lower bound $W(S_2,q)_\ell$ compares with the
exact $W(sq,q)$ and $W(sc,q)$ is to compare the large-$q$ series expansions for
these three functions.  For this purpose, it is first appropriate to give some
relevant background on large-$q$ series expansions for $W(\{G\},q)$ functions.
Since there are $q^n$ possible colorings of the vertices of an $n$-vertex graph
$G$ with $q$ colors if no conditions are imposed, an obvious upper bound on the
number of proper $q$-colorings of the vertices of $G$ is $P(G,q) \le q^n$.
This yields the corresponding upper bound $W(\{G\},q) < q$.  Hence, it is
natural to define a reduced function that has a finite limit as $q \to \infty$,
\beq
W_r(\{G\},q) = q^{-1}W(\{G\},q) \ . 
\label{wr}
\eeq
For a lattice or, more generally, a graph whose vertices have bounded degree,
$W_r(\{G\},q)$ is analytic about $1/q=0$. ($W_r(\{G\},q)$ is non-analytic at
$1/q=0$ for certain families of graphs that contain one or more vertices with
unbounded degree as $n \to \infty$, although the presence of a vertex with
unbounded degree in this limit does not necessarily imply non-analyticity of
$W_r(\{G\},q)$ at $1/q=0$ \cite{wa,sb}.)  It is conventional to express the
large-$q$ Taylor series for a function that has some factors removed from
$W_r$, since this function yields a simpler expansion.  A chromatic polynomial
has the general form
\beq
P(G,q) = \sum_{j=0}^{n-k(G)} (-1)^j a_{n-j} q^{n-j} \ , 
\label{p}
\eeq
where the $a_{n-j} > 0$ and $k(G)$ is the number of connected components of $G$
(taken here to be $k(G)=1$ without loss of generality).  One has $a_n = 1$,
$a_{n-1} = e(G)$, and, provided that the girth $g(G) > 3$ \cite{g}, as is the
case here, $a_{n-2} = {e(G) \choose 2}$.  A $\kappa$-regular graph is a graph
such that each vertex has degree (coordination number) $\kappa$.  For a
$\kappa$-regular graph, $e(G) = \kappa n/2$.  The coefficients of the three
terms of highest degree in $q$ in $P(G,q)$ for a $\kappa$-regular graph are
precisely the terms that would result from the expansion of
$[q(1-q^{-1})^{\kappa/2}]^n$.  Hence, for a $\kappa$-regular graph or lattice,
one usually displays the large-$q$ series expansions for the reduced function
\beq
\overline W(\Lambda,q) = \frac{W(\Lambda,q)}{q(1-q^{-1})^{\kappa/2}} \ . 
\label{wbar}
\eeq
The large-$q$ Taylor series for this function can be written in the form
\beq
\overline W(\Lambda,q)=1+\sum_{j=1}^\infty w_{\Lambda,j} y^j \ , 
\label{wseries}
\eeq
where 
\beq
y = \frac{1}{q-1} \ . 
\label{y}
\eeq

 The two results that we shall need here are the large-$q$ (i.e., small-$y$)
Taylor series for $\overline W(sq,q)$ and $\overline W(sc,q)$.  The large-$q$
series for $\overline W(sq,q)$ was calculated to successively higher orders
in \cite{nagle71}-\cite{bk94}.  Here we only quote the terms to $O(y^{11})$:
\beqs
\overline W(sq,q) & = & 1 + y^3 + y^7 +3y^8 + 4y^9 + 3y^{10} \cr\cr
                  & + & 3y^{11} + O(y^{12}) \ .
\label{wsqseries}
\eeqs

As noted above, lower bounds on $W(\Lambda,q)$ obtained from the inequality
(\ref{london}) for two-dimensional lattices $\Lambda$ were found to be quite
close to the actual values of the respective $W(\Lambda,q)$ for a large range
of values of $q$.  This can be understood for large values of $q$ from the fact
that they coincide with the large-$q$ expansions to many orders, and the
agreement actually extends to values of $q$ only moderately above $q=2$. For
example, the lower bound on $W(sq,q)$ in \cite{biggs77} is equivalent to
$\overline W(sq,q) \ge (1+y^3)$. This agrees with the small-$y$ series up to
order $O(y^6)$, as is evident from comparison with Eq. (\ref{wsqseries}). This
lower bound also agrees quite closely with the value of $W(sq,q)$ determined by
Monte Carlo simulations in \cite{w,ww,w3} (see Table 1 of \cite{w} and Table 1
of \cite{ww}).  We include this comparison here in Table \ref{wvalues}.  For
our purposes, it is sufficient to quote the results from Ref. \cite{w} only to
three significant figures. Since we are using large-$q$ series for this
comparison, we list the results in Table \ref{wvalues} for a set of values $q
\ge 4$.  As another example, the lower bound obtained for the honeycomb lattice
in Ref. \cite{w3}, $\overline W(hc,q) \ge (1+y^5)^{1/2}$, agreed with the
small-$y$ series for $\overline W(hc,q)$ to $O(y^{10})$.  Thus, it was found
that for all of the cases studied, $W(\Lambda,q)_\ell$ provides not only a
lower bound on $W(\Lambda,q)$, but a rather good approximation to the latter
function.  It is thus reasonable to expect that this will also be true for the
lower bounds $W(S_{(L_z)_{BC_z}},q)_\ell$ for the slabs $S_{L_z}$ of the simple
cubic lattice considered here, of infinite extent in the $x$ and $y$ directions
and of thickness $L_z$ in the $z$ direction.

From ingredients given in Ref. \cite{baker71}, we have calculated a large-$q$
expansion of the $\overline W(sc,q)$ for the simple cubic ($sc$) lattice and
obtain 
\beq
\overline W(sc,q) = 1 + 3y^3 + 22y^5 + 31y^6 + O(y^7) \ . 
\label{wbarsc_smally}
\eeq
In Table \ref{wvalues} we list the corresponding values of $W(sc,q)$ obtained
from this large-$q$ series, denoted $W(sc,q)_{ser.}$, for $q \ge 4$.  We also
list estimates of $W(sc,q)$, denoted $W(sc,q)_{MC}$, for $4 \le q \le 6$ from
the Monte Carlo calculations in Ref. \cite{cp88}.  One sees that the
approximate values obtained from the large-$q$ series are close to the
estimates from Monte Carlo simulations even for $q$ values as low as $q=4$.

The coordination number for the $S_{2_F}$ slab of the
simple cubic lattice (of infinite extent in the $x$ and $y$ directions) is
$\kappa(S_{2_F})=5$. We thus analyze the reduced function 
$\overline W(S_2,q)_\ell = W(S_2,q)_\ell/[q(1-q^{-1})^{5/2}]$. 
This has the large-$q$ (small-$y$) expansion
\beq
\overline W(S_2,q)_\ell = 1 + 2y^3 + 2y^5 + 9y^6 + O(y^7) \ . 
\label{wbars2Fseries}
\eeq
As this shows, $\overline W(S_2,q)_\ell$ provides an interpolation between
$\overline W(sq,q)$ and $\overline W(sc,q)$; for example, the coefficient of
the $y^3$ term is 1 for $\Lambda=sq$, 2 for $\Lambda=S_2$, and 3 for
$\Lambda=sc$.  Furthermore, the coefficient of the $y^5$ term is 0 for
$\Lambda=sq$, 2 for $\Lambda=S_2$, and 22 for $\Lambda=sc$.  This is in
agreement with the fact that the exact functions $W(S_{(L_z)_F},q)$ 
interpolate between $W(sc,q)$ and $W(sc,q)$ as $L_z$ increases from 1 to
$\infty$ \cite{w2d} and the expectation, as discussed above, that 
$W(S_{(L_z)_F},q)_\ell$ should be close to $W(S_{(L_z)_F},q)$.

\section{Results for Slabs of Thickness $L_z=3, \ 4$ with FBC$_z$}

For the slab of the simple cubic lattice with thickness $L_z=3$ and FBC$_z$,
denoted $S_{3_F}$, the transverse slice is the graph $sq[3_F \times
(L_y)_{BCy}]$.  The chromatic polynomials $P(sq[3_F \times (L_y)_F],q)$,
$P(sq[3_F \times (L_y)_P],q)$, and $P(sq[3_F \times (L_y)_{TP}],q)$ (where $TP$
denotes twisted periodic, i.e., M\"obius BC) were computed for arbitrary $L_y$
in Refs. \cite{strip}, \cite{wcyl}, and \cite{pm}, respectively, and the
maximal $\lambda$ was shown to be the same for all of these boundary
conditions.  For the reader's convenience, we list this
$\lambda_{ts(3_F),max}$ in Eqs. (\ref{lamstrip3}) and (\ref{r3}) of the
Appendix.  The other input that is needed to obtain the lower bound in
Eq. (\ref{wlowbound}) is the maximal $\lambda$ for the chromatic polynomial of
the $sc[2_F \times 3_F \times L_y]$ tube graph, i.e.,
$\lambda_{tube(3_F),max}$.  The relevant transfer matrix that determines the
chromatic polynomial for this tube graph was given with Ref. \cite{hd}.
Because it is $13 \times 13$ dimensional, one cannot solve the corresponding
characteristic polynomial analytically to obtain $\lambda_{tube(3_F),max}$ for
general $q$.  However, one can calculate $\lambda_{tube(3_F),max}$
numerically, and we have done this.  Combining these results with
Eqs. (\ref{lamstrip3}) and (\ref{r3}), we then evaluate the lower bound
$W(S_{3_F},q)_\ell$ by evaluating the $L_z=3$ special case of
(\ref{wlowbound}).

For the slab of the simple cubic lattice with thickness $L_z=4$ and FBC$_z$,
$S_{4_F}$, the transverse slice is the graph $sq[4_F \times (L_y)_{BCy}]$.
Here the maximal $\lambda_{ts(4_F),max}$ is the solution of the cubic equation
(\ref{strip4_gden}) given in the Appendix.  One also needs the maximal
$\lambda$ for the chromatic polynomial of the $sc[2_F \times 4_F \times L_y]$
tube graph, i.e., $\lambda_{tube(4_F),max}$.  The relevant ($136 \times 136$
dimensional) transfer matrix for this tube graph was calculated for
Ref. \cite{hd}, and we have used this to compute $\lambda_{tube(4_F),max}$
numerically. We then obtain the lower bound $W(S_{4_F},q)_\ell$ from the
$L_z=4$ special case of Eq. (\ref{wlowbound}).  The results for $W(S_{3_F},q)$
and $W(S_{4_F},q)$ are listed in Table \ref{wvalues}.

\section{Result for Slab of Thickness $L_z=3$ with PBC$_z$}

It is also of interest to obtain a lower bound for $W$ for a slab with periodic
boundary conditions in the $z$ direction, since these minimize finite-volume
effects.  For this purpose we consider the slab of the simple cubic lattice 
with thickness $L_z=3$ and PBC$_z$, $S_{3_P}$.  In this case the 
transverse slice is the graph $sq[3_P \times (L_y)_{BCy}]$. For FBC$_y$ the
chromatic polynomial involves only one $\lambda$, and this is also the maximal
$\lambda$ for PBC$_y$ and TPBC$_y$ \cite{tk}, viz., 
\beq
\lambda_{ts(3_P),max} = q^3-6q^2+14q-13 \ . 
\label{lam3pdom}
\eeq
One then needs $\lambda_{tube(3_P),max}$.  The relevant ($4 \times 4$
dimensional) transfer matrix for this tube graph was calculated for
Ref. \cite{hd}, and we have used this to compute $\lambda_{tube(3_P),max}$
numerically. The results for $W(S_{3_P},q)$ are given in Table \ref{wvalues}.

\section{Discussion}

Since the slabs of infinite extent in the $x$ and $y$ directions and of finite
thickness $L_z$ geometrically interpolate between the square and simple cubic
lattices, it follows that the resultant $W$ functions for these slabs
interpolate between $W(sq,q)$ and $W(sc,q)$ \cite{w2d}.  Given that it was
shown previously that the lower bounds $W(\Lambda,q)_\ell$ obtained by the
coloring matrix method are quite close to the actual values of the respective
$W(\Lambda,q)$ for a number of two-dimensional lattices, this is also expected
to be true for the $W(S_{(L_z)_{BC_z}},q)_\ell$ bounds.  We have shown above
how $W(S_2,q)_\ell$ interpolates between $W(sq,q)$ and $W(sc,q)$ via a
comparison of the large-$q$ series expansions for these three functions.  Table
\ref{wvalues} provides a further numerical comparison for $W(S_{3_F},q)_\ell$,
$W(S_{4_F},q)$, and $W(S_{3_P},q)$ with $W(sq,q)$ and $W(sc,q)$, the latter
being determined to reasonably good accuracy from large-$q$ series expansions
and, where available, Monte Carlo measurements. As noted, the lower end of the
range of $q$ values for the comparison is chosen as $q=4$ in view of the use of
large-$q$ series.

For sections of lattices, and, more generally, graphs that are not
$\kappa$-regular, one can define an effective vertex degree (coordination
number) as \cite{wn} 
\beq
\kappa_{eff} = \frac{2e(G)}{n(G)} \ . 
\label{kappaeff}
\eeq
For $3 \le L_z < \infty$, the slab of the simple cubic lattice (of infinite
extent in the $x$ and $y$ directions) with FBC$_z$ is not $\kappa$-regular, 
but has the effective coordination number
\beq
\kappa_{eff}(S_{(L_z)_F}) = 2 \bigg ( 3 - \frac{1}{L_z} \bigg ) \ . 
\label{kappas}
\eeq
We observe that for the $q$ values considered in Table \ref{wvalues}, $W(sq,q)
> W(S_2,q)_\ell > W(S_{3_F},q)_\ell > W(S_{4_F},q) > W(sc,q)$.  The fact that
for fixed $q$, the exact function $W(S_{(L_z)_F},q)$ is a non-increasing
function of $L_z$, and, for $q > 2$, a monotonically decreasing function of
$L_z$, follows from a theorem proved in Ref. \cite{w2d}. To the extent that the
lower bounds $W(S_{(L_z)_F},q)_\ell$ lie close to the actual values of
$W(S_{(L_z)_F},q)$, it is understandable that they also exhibit the same strict
monotonicity.  As was noted in Ref. \cite{w2d}, the reason for the monotonicity
of the exact values is that the number of proper $q$-colorings per vertex of a
lattice graph is more highly constrained as one increases the effective
coordination number of the lattice section. (This is also evident in Fig. 5 of
\cite{w}.)  In the present case, the monotonicity can be seen as a result of
the fact that the effective coordination number increases monotonically as a
function of $L_z$.  

The use of periodic boundary conditions in the $z$ direction minimizes
finite-size effects, so that for a given $L_z$, $W(S_{(L_z)_P},q)$ would be
expected to be closer to $W(sc,q)$ than $W(S_{(L_z)_F},q)$ \cite{w2d}.  Again,
to the extent that the lower bounds are close to the actual $W$ functions for
these respective slabs, one would expect $W(S_{(L_z)_P},q)_\ell$ to be closer
than $W(S_{(L_z)_F},q)_\ell$ to $W(sc,q)$.  Our results agree with this
expectation. In contrast to $W(S_{(L_z)_F},q)$, $W(S_{(L_z)_P},q)$ is not, in
general, a non-increasing function of $L_z$, as was discussed in general in
\cite{w2d} (see Fig. 1 therein).  Thus, values of $W(S_{(L_z)_P},q)$, and
hence, {\it a fortiori}, $W(S_{(L_z)_P},q)_\ell$, may actually lie slightly
below those for $W(sc,q)$, as is evident for the $W(S_{3_P},q)_\ell$ entries in
Table \ref{wvalues}.

\section{Conclusions}

In this paper we have calculated rigorous lower bounds for the ground state
degeneracy per site $W$, equivalent to the ground state entropy $S_0=k_B\ln W$,
of the $q$-state Potts antiferromagnet on slabs of the simple cubic lattice
that are infinite in two directions and finite in the third. Via comparison
with large-$q$ expansions and numerical evaluations, we have shown how the
results interpolate between the square (sq) and simple cubic (sc) lattices.

\bigskip
\bigskip

Acknowedgments: This research was supported in part by the NSF grant
PHY-06-53342. RS expresses his gratitude to coauthors on previous related
works, in particular, S.-H. Tsai and J. Salas, as well as N. Biggs, 
S.-C. Chang, and M. Ro\v{c}ek.

\section{Appendix} 

We note the following results on ${\mathbb E}^d$ lattices and lattice sections:
$W(\Lambda_{bip.},2)=1$ for any bipartite lattice; $W(sq,3)=(4/3)^{3/2}$
\cite{lieb67}; and $W(\{L\},q)=W(\{C\},q)=q-1$, where $L_n$ and $C_n$ denote
the $n$-vertex line and circuit graphs. For the infinite-length square-lattice
strip of width 2, $W(sq[2_F \times \infty],q) = W(sq[2_P \times
\infty],q)=\sqrt{q^2-3q+3}$, where, as in the text, the subscripts $F$ and $P$
denote free and periodic boundary conditions in the direction in which the
strip is finite.  For the infinite-length strip of the square lattice with
(transverse) width 3 and free transverse boundary conditions, $sq[3_F \times
\infty]$ \cite{strip,wcyl,pm}
\beq
W(sq[3_F \times \infty],q) = (\lambda_{3_F,max})^{1/3}
\label{w3m}
\eeq
where
\beq
\lambda_{3_F,max} = \frac{1}{2} 
\bigg [(q-2)(q^2-3q+5) + \sqrt{R_3} \ \bigg ]
\label{lamstrip3}
\eeq
with
\beq
R_3 = (q^2-5q+7)(q^4-5q^3+11q^2-12q+8) \ . 
\label{r3}
\eeq

For the infinite-length strip of the square lattice with width 4 and free
transverse boundary conditions, $sq[4_F \times \infty]$ \cite{strip,s4}
\beq
W(sq[4_F \times \infty],q) = (\lambda_{4_F,max})^{1/4}
\label{w4m}
\eeq
where $\lambda_{4_F,max}$ is the largest root of the cubic equation
\beq
x^3 + b_{4_F,1} x^2 + b_{4_F,2} x + b_{4_F,3} = 0
\label{strip4_gden}
\eeq
with
\beq
b_{4_F,1} = -q^4+7q^3-23q^2+41q-33
\label{strip4_b1}
\eeq
\beqs
b_{4_F,2} & = & 2q^6-23q^5+116q^4-329q^3+553q^2-517q+207 \cr\cr
& &
\label{strip4_b2}
\eeqs
and
\beqs
b_{4_F,3} & = & -q^8+16q^7-112q^6+449q^5-1130q^4+1829q^3 \cr\cr
    & - & 1858q^2+1084q-279 \ .
\label{strip4_b3}
\eeqs

\newpage

\begin{widetext}
\begin{table}
\caption{\footnotesize{Comparison of lower bounds
$W(S_{(L_z)_{BC_z}},q)_{\ell}$ for $(L_z)_{BCz}=2_F=2_P, \ 3_F, \ 4_F, \ 3_P$
with approximate values of $W(\Lambda,q)$ for the square (sq) and simple cubic
(sc) lattices $\Lambda$, as determined from large-$q$ series expansions,
denoted $W(\Lambda,q)_{ser.}$ and, where available, Monte Carlo simulations, denoted
$W(\Lambda,q)_{MC}$. We also list $W(sq,q)_{\ell}$ for reference. See text for
further details.}}
\begin{center}
\begin{tabular}{cccccccccc}
$q$ & $W(sq,q)_{MC}$ & $W(sq,q)_{ser.}$ & $W(sq,q)_\ell$ & 
$W(S_{2_F},q)_\ell$ & $W(S_{3_F},q)_\ell$ & $W(S_{4_F},q)_\ell$ 
& $W(sc,q)_{ser.}$ & $W(sc,q)_{MC}$ & $W(S_{3_P},q)_\ell$ \\ 
\hline
 4 & 2.34  & 2.34  & 2.33  & 2.13 & 2.07  & 2.04 & 2.06 & 1.9 & 1.78  \\
 5 & 3.25  & 3.25  & 3.25  & 2.96 & 2.875 & 2.83 & 2.75 & 2.7 & 2.62  \\
 6 & 4.20  & 4.20  & 4.20  & 3.87 & 3.765 & 3.71 & 3.58 & 3.6 & 3.51  \\
 7 & 5.17  & 5.17  & 5.17  & 4.81 & 4.69  & 4.64 & 4.48 & $-$ & 4.43  \\
 8 & 6.14  & 6.14  & 6.14  & 5.76 & 5.64  & 5.58 & 5.41 & $-$ & 5.37  \\
 9 & 7.125 & 7.125 & 7.125 & 6.73 & 6.605 & 6.54 & 6.36 & $-$ & 6.325 \\ 
10 & 8.11  & 8.11  & 8.11  & 7.71 & 7.58  & 7.51 & 7.32 & $-$ & 7.29  \\ 
\hline
100 & $-$  & 98.0  & 98.0  & 97.5 & 97.4  & 97.3 & 97.0 & $-$ & 97.0  \\
\hline
\end{tabular}
\end{center}
\label{wvalues}
\end{table}
\end{widetext}


\begin{thebibliography}{99}
\bibitem{frus}
This is in contrast to systems where nonzero ground state entropy is 
associated with frustration, including the Ising antiferromagnet on the
triangular lattice, spin glasses, and spin ice materials such as pyrochlores.
\bibitem{g}
The girth of a graph $G$, $g(G)$ is defined as the number of edges of a
minimal-length closed circuit on $G$.
\end{thebibliography}
\end{document}